\documentclass{iau} 
\usepackage{graphicx}

\title[Integrated-light spectroscopy of globular clusters] %% give here short title %%
{Modeling and analysis of medium-\\ resolution integrated-light spectra of \\ 
globular clusters  in dwarf galaxies}

\author[Margarita Sharina \& Vladislav Shimansky]   %% give here short author list %%
{Margarita E. Sharina$^1$
 \and Vladislav V. Shimansky$^2$}

\affiliation{$^1$Special Astrophysical Observatory,
Russian Academy of Sciences, \\ Nizhnii Arkhyz, 369167, Russia,  email: 
{\tt sme@sao.ru} \\[\affilskip]
$^2$ Kazan (Volga region) Federal University, Kazan, 420008 Russia
 \\email: {\tt Slava.Shimansky@kpfu.ru}}

\pubyear{2019}
\volume{351}  
\jname{Star Clusters: From the Milky Way to the Early Universe}
\editors{A. Bragaglia, M.B. Davies, A. Sills \& E. Vesperini, eds.}
\begin{document}

\maketitle
%. CONTINUE EDITING FROM HERE

\begin{abstract}
The study of ages, helium mass fraction (Y) and chemical composition of globular clusters in dwarf galaxies is important for understanding the physical conditions at the main evolutionary stages of the host galaxies and for constraining the build-up histories of large galaxies. We present the analysis of integrated-light spectra of 8 extragalactic and 20 Galactic globular clusters (GCs) using our population synthesis method. We calculate synthetic spectra of GCs according to the defined stellar mass functions using model atmospheres and stellar parameters ($[Fe/H]$, $T_{eff}$, and $log g$) set by theoretical isochrones. The main advantage of our method is the ability to determine not only chemical composition but also the age and mean Y in a cluster by modelling and analysis of Balmer absorption lines. The knowledge of Y and anomalies of light elements in star clusters is one of the key points for understanding the phenomenon of multiple stellar populations.
\end{abstract}

\firstsection % if your document starts with a section,
%\section{Introduction}
%\cite[Anders \& Zinner (1993)]{AndersZinner93} and 
%\cite[Ott (1993)]{Ott93}.
%Fig.\,\ref{fig1}
%{\underline{\it About our method}}.
\section{About our method}
We use integrated-light spectra of globular clusters (GCs) and the method of 
model stellar atmospheres to derive ages, Y and chemical composition of the
studied objects (\cite[Sharina \etal\ 2017]{Sharina18} and references therein).
Synthetic integrated-light spectra calculation is based on the plane-parallel, hydrostatic
stellar atmosphere models by \cite[Castelli \& Kurucz (2003)]{CastelliKurucz03}. 
The lists of atomic and molecular lines are taken from the R.L. Kurucz web site (http://kurucz.harvard.edu/linelists.html).
%Synthetic spectra of stars in star clusters are calculated according to their parameters: $[Fe/H]$, $T_{eff}$, and $log g$. 
In this work we use stellar evolutionary isochrones by \cite[Bertelli \etal\ (2008)]{Bertelli08}. 
The calculated synthetic spectra of
individual stars are summed according to the mass function by \cite[Chabrier (2005)]{Chabrier05}.
%The lists of atomic and molecular lines are taken from the R.L. Kurucz web site(http://kurucz.harvard.edu/linelists.html).
Comparison of the shapes and intensities of the observed and the model Balmer
line profiles allows us to derive
 the age, Y, and horizontal branch (HB) type of a GC.
 The influence of Y and age on the H lines is not equivalent. 
The temperatures of main-sequence turnoff stars become higher with the decreasing age. 
This means that the depths of the cores and wings of the Balmer lines simultaneously strengthen. 
Increasing of Y results in higher luminosities of hot HB stars 
and in the increasing of the depth of the cores and wings of the Balmer lines in their spectra.
In hotter stellar atmospheres pressure broadening and H-continuum opacity non-synchronously diminish. 
%For HB stars hotter than $T_{eff}\sim9000$~K ionization effects play a role.
As a result, the depths of the wings and cores of H$_{\delta}$, H$_{\gamma}$, and H$_{\beta}$ change differently 
with the change of Y, because hot HB stars contribute mainly the blue part of the spectrum.
\begin{figure}[t]
% \vspace*{-2.0 cm}
\begin{center}
 \includegraphics[width=4.5in]{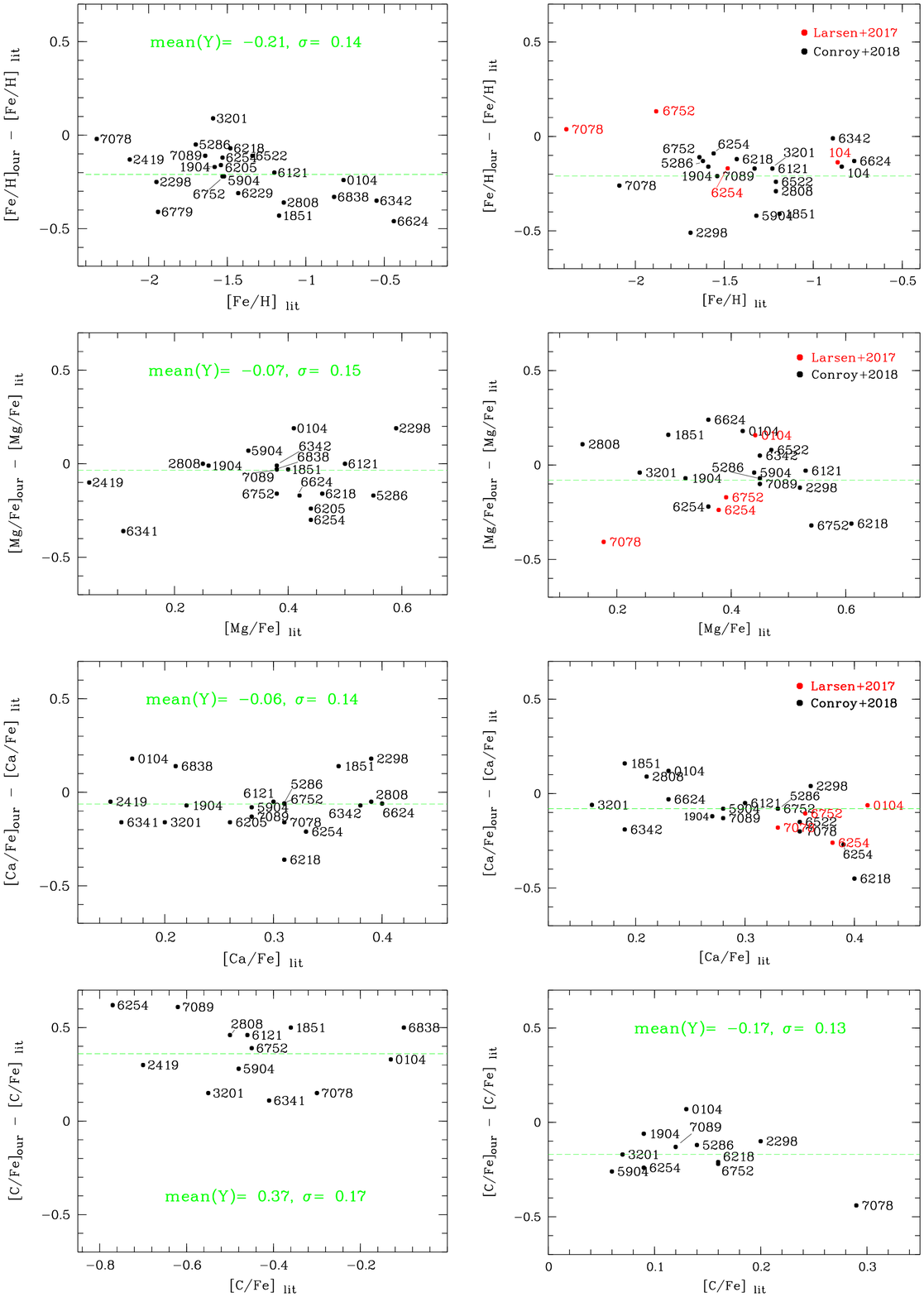} 
% \vspace*{-4.0 cm}
 \caption{Comparison of the abundances of chemical elements determined using our method 
 with the abundances from high-resolution spectroscopic
studies of red giants in the clusters (left) (\cite[Pritzl \etal\ 2005]{Pritzl05}, \cite[Roediger \etal\ 2014]{Roediger14}) 
and with the abundances
determined using integrated light spectra of GCs (right) (\cite[Larsen \etal\ 2017]{Larsen17}, 
\cite[Conroy \etal\ 2018]{Conroy18}). NGC numbers are indicated.}
   \label{fig1}
\end{center}
\end{figure}
\begin{table}
  \begin{center}
  \caption{Logarithmic age in years, helium mass fraction and metallicity in dex for GCs in M33 
  (CBF~98 and CBF~28), for nuclear GCs in dSphs ESO269-66 and KKs3, and for GCs in the  M31 neighbourhood 
  ([SD09]~GC7, MayallII, MGC~1 and Bol~298). Results derived using the described method are highlighted 
  in bold.}
  \label{tab1}
 {\scriptsize
  \begin{tabular}{|l|c|c|c|}\hline 
Object   & $log(T)$ & $Y$ & $[Fe/H]$  \\ \hline 
{\bf CBF 98}& {\bf 10.05$\pm$0.05 $^0$}& {\bf 0.30 $^0$} & {\bf -1.40$\pm$0.15 $^0$} \\ 
           & 10.04$\pm$0.05 $^{1}$       & --         & -1.30$\pm$0.10 $^{1}$  \\     
{\bf CBF 28}& {\bf 10.0$\pm$0.05 $^0$}& {\bf 0.23 $^0$} & {\bf -1.55$\pm$0.15 $^0$}  \\ 
{\bf KKs3 GC1} &  {\bf 10.1$\pm$0.1 $^2$} &  {\bf 0.30$\pm$0.03 $^2$}& {\bf-1.5$\pm$0.2 $^2$} \\ 
{\bf ESO269-66 GC1}& {\bf 10.1$\pm$0.1 $^2$} & {\bf 0.30$\pm$0.03 $^2$}& {\bf -1.5$\pm$0.2 $^2$} \\ 
{\bf SD09 GC7}& {\bf 10.0$\pm$0.1 $^3$} & {\bf 0.30$\pm$0.03 $^3$} & {\bf -1.8$\pm$0.1 $^3$} \\  
         & 9.9$\pm$0.12 $^4$ &    --        & -1.5$\pm$0.2 $^4$, -1.8$\pm$0.3  $^5$ \\ 
{\bf MayallII} & {\bf 10.15$\pm$0.05 $^3$}& {\bf 0.26$\pm$0.02 $^3$}& {\bf -1.00$\pm$0.05 $^3$} \\ 
         & 10.18 $^6$,  10.08  $^7$      &    --        & -0.95$\pm$0.09  $^6$, -1.08$\pm$0.09 $^8$\\ 
{\bf MGC~1}   & {\bf 10.0$\pm$0.05 $^3$} & {\bf 0.30$\pm$0.03 $^3$}& {\bf -2.20$\pm$0.1 $^3$}  \\
         & 9.7$\pm$0.1 $^9$  &    --        &  -2.14 $^{10}$       \\      
{\bf Bol~298}  & {\bf 10.0$\pm$0.1 $^3$}  & {\bf 0.30$\pm$0.03 $^3$}& {\bf -1.85$\pm$0.1 $^3$} \\ 
         & 10.3$\pm$0.1 $^9$, 10.13 $^{11}$    &    --        & -2.14 $^{10}$, -2.07$\pm$0.18 $^{11}$ \\ 
 \hline
  \end{tabular}
  }
 \end{center}
\vspace{1mm}
 \scriptsize{
 %\begin{tabular}
 % {lp{80mm}}
{\it Notes:}
{\bf $^0$this work}; $^{1}$Sharina et al. 2010; {\bf $^2$Sharina et al. 2017; $^3$Sharina et al. 2018;}  
$^4$Sharina \& Davoust \\ 2009; $^5$Veljanoski et al. 2013; $^6$Meylan et al. 2001;  $^7$Ma 2009;
$^8$Huchra et al. 1991;  $^9$Ma 2012; $^{10}$Mackey et al. 2007; $^{11}$Fan et al. 2011.}
\end{table}

\section{Results}

\begin{figure}[h]
\vspace*{-0.3cm}
%\begin{center}
\begin{tabular}{p{0.32\textwidth}p{0.32\textwidth}p{0.32\textwidth}}
%\hspace{0.1cm}
 \vspace{0.4cm}
  \includegraphics[width=1.75in,angle=-90]{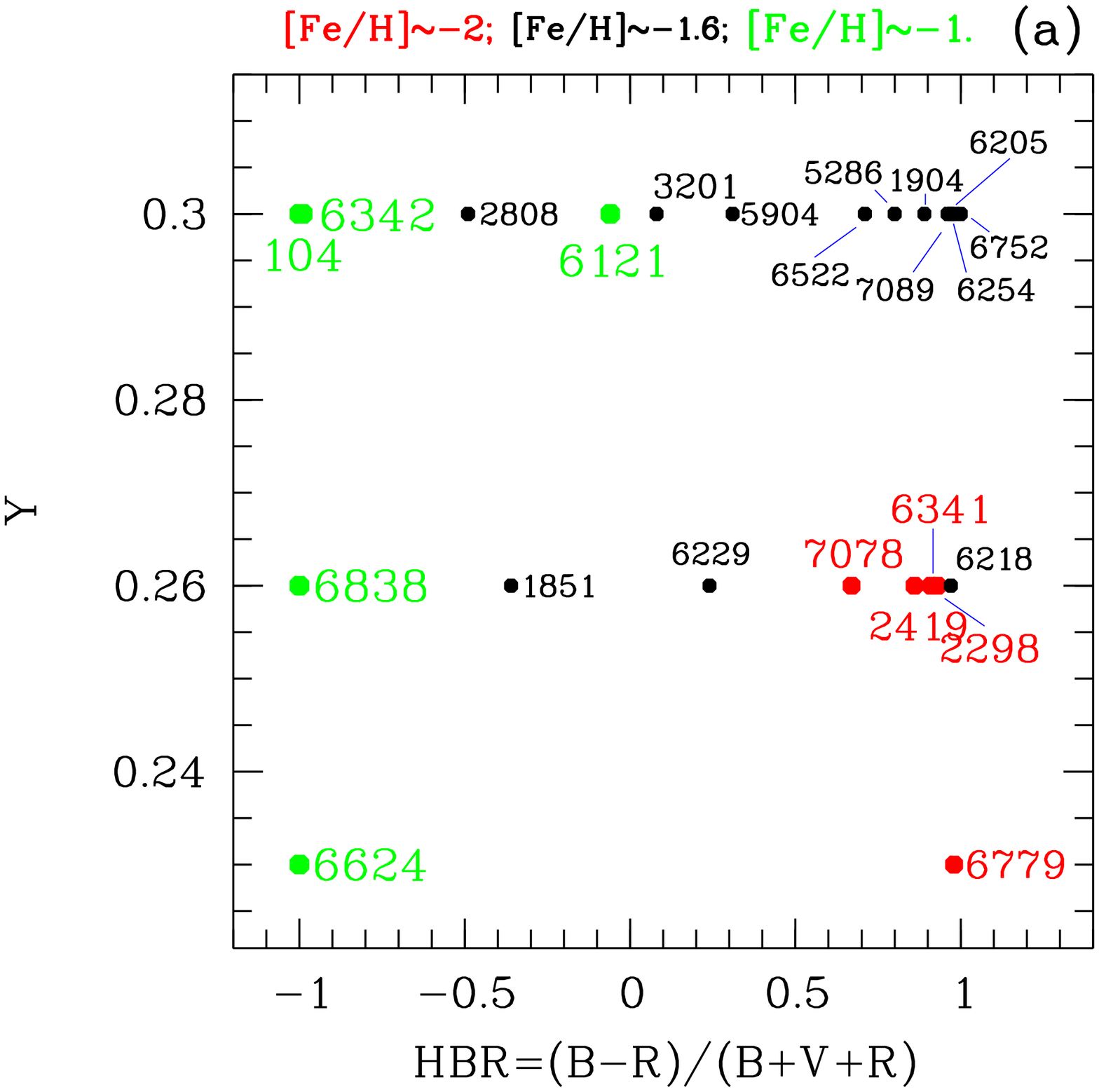} &
%\hspace{0.1cm}
 \vspace{0.4cm} 
 \includegraphics[width=1.75in,angle=-90]{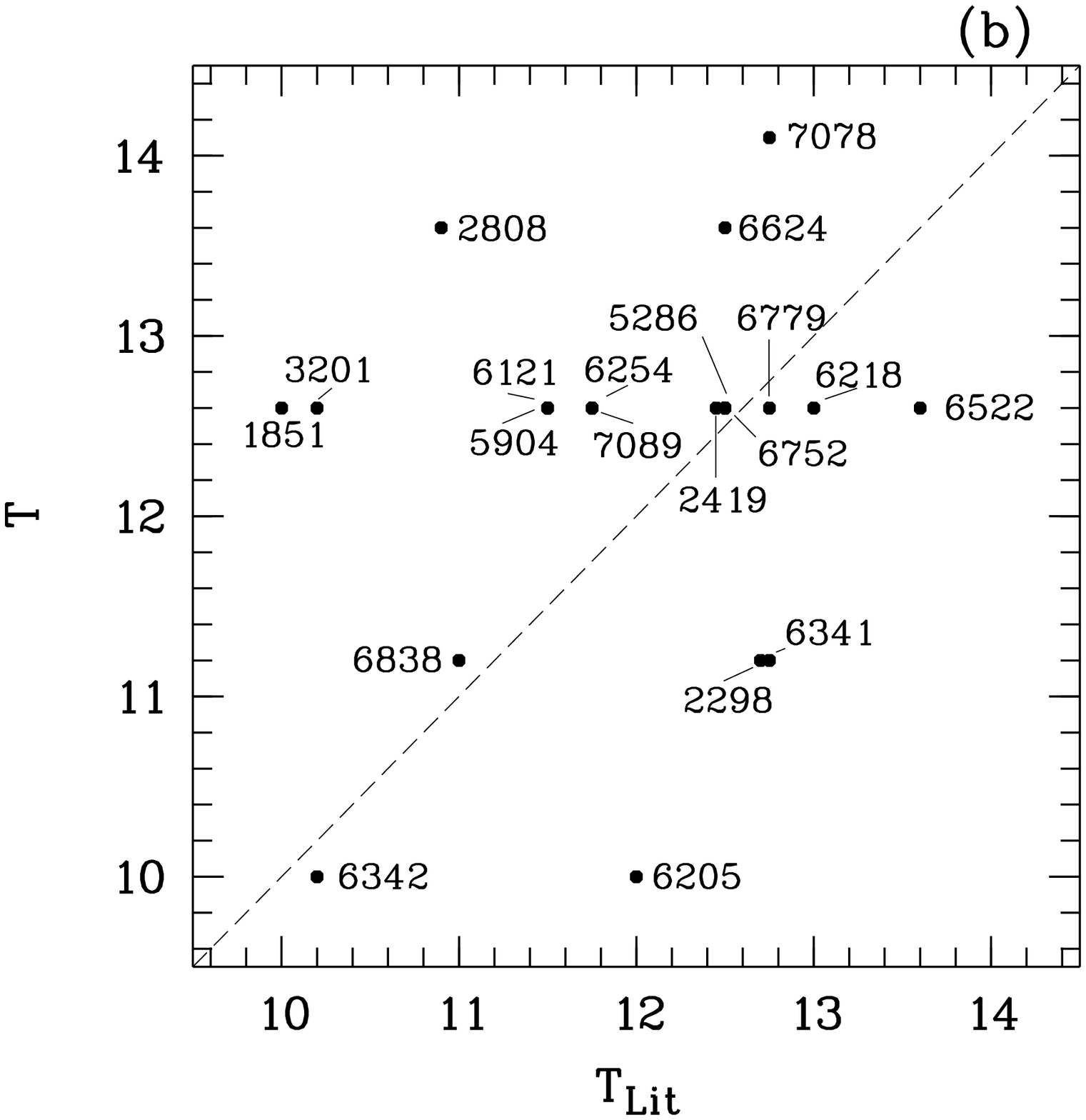} &
%  \hspace{-0.2cm}
\includegraphics[width=1.75in,angle=-180]{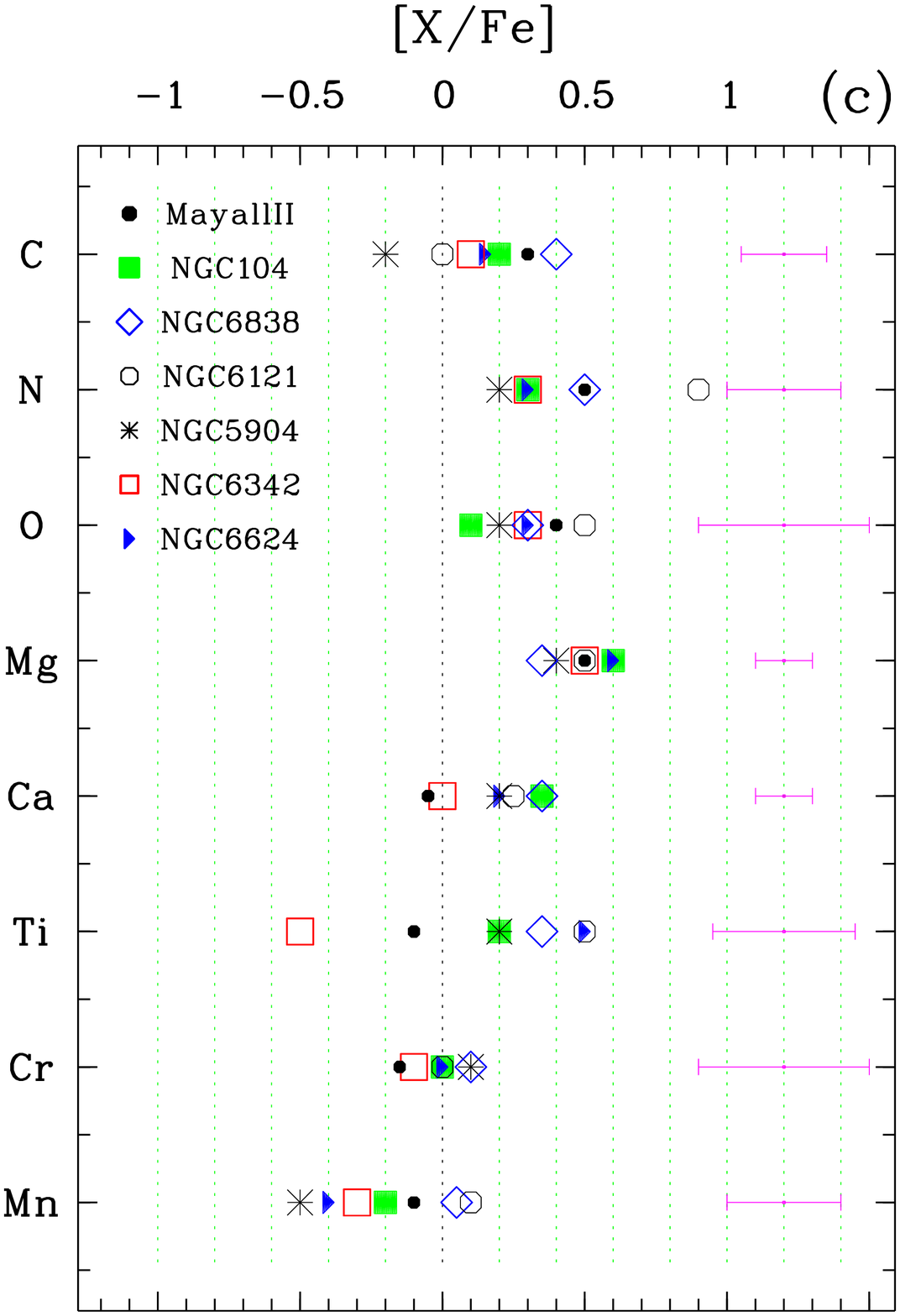} \\
\end{tabular}
% \vspace*{-4.0 cm}
 \caption{(a) Comparison of the Y values derived using our 
 method with the Horizontal-branch ratios (\cite[Harris, 1996]{Harris96});
 (b) comparison of the ages derived using our method with
 the literature values listed by \cite[Roediger \etal\ (2014)]{Roediger14};
 (c) abundances of chemical elements determined using our method for MayallII in  M31 
 and for 6 Galactic GCs of similar metallicity. Typical errors of elemental abundance 
 determination with our method are shown.}
\label{fig2}
%\end{center}
\end{figure}

We use high signal-to-noise ($\rm S/N\!\sim\!100$) medium-resolution 
($\rm FWHM\!\sim\!3\!-\!5$\AA) integrated-light spectra of globular clusters in a wide spectral range
(3900-5500\AA) (e.g. \cite[Sharina \etal\ 2017]{Sharina17}). 
The method is tested using spectra of Galactic GCs observed  with the CARELEC spectrograph at the 1.93-m telescope 
of the Haute-Provence observatory and the spectra from \cite[Schiavon \etal\ (2005)]{Schiavon05}.
The results are shown in Fig.\,\ref{fig1} and Fig.\,\ref{fig2} (panels a and b). 
It can be seen that
%the mean standard deviations of the differences between our and literature values are 
%$\sigma\!\sim0.15$ for Fe, Mg, Ca and  $\sigma\!\sim0.17$ for C.
there is a systematic difference $\sim0.36$~dex between the carbon abundances determined using our method 
and those from high-resolution spectroscopic
studies (left panel of Fig.\,\ref{fig1}). Note that this difference disappears if we compare our measurements of 
$[C/Fe]$ with the corresponding values determined using integrated-light spectra (right panel of Fig.\,\ref{fig1}).
We suggest that this difference is the result of the stellar evolution (\cite[Sharina \etal\ 2017]{Sharina17}). 
%and mixing of the CNO cycle material within the atmospheres of RGB stars (\cite[Gratton \etal\ 2012]{Gratton12}). 
Fig.\,\ref{fig2} (panels a and b) shows that most of the clusters with high Y have blue horizontal branches 
and that for most of the studied objects the estimated ages agree with the literature values within $\sim 1$~Gyr.

\begin{figure}[]
% \vspace*{-2.0 cm}
\begin{center}
\begin{tabular}{p{0.325\textwidth}p{0.325\textwidth}p{0.325\textwidth}}
% \hspace{-0.39cm}
% \includegraphics[width=1.4in,angle=-180]{MayallIIelem.ps} &
 \hspace{-0.4cm}
  \includegraphics[width=1.85in,angle=-180]{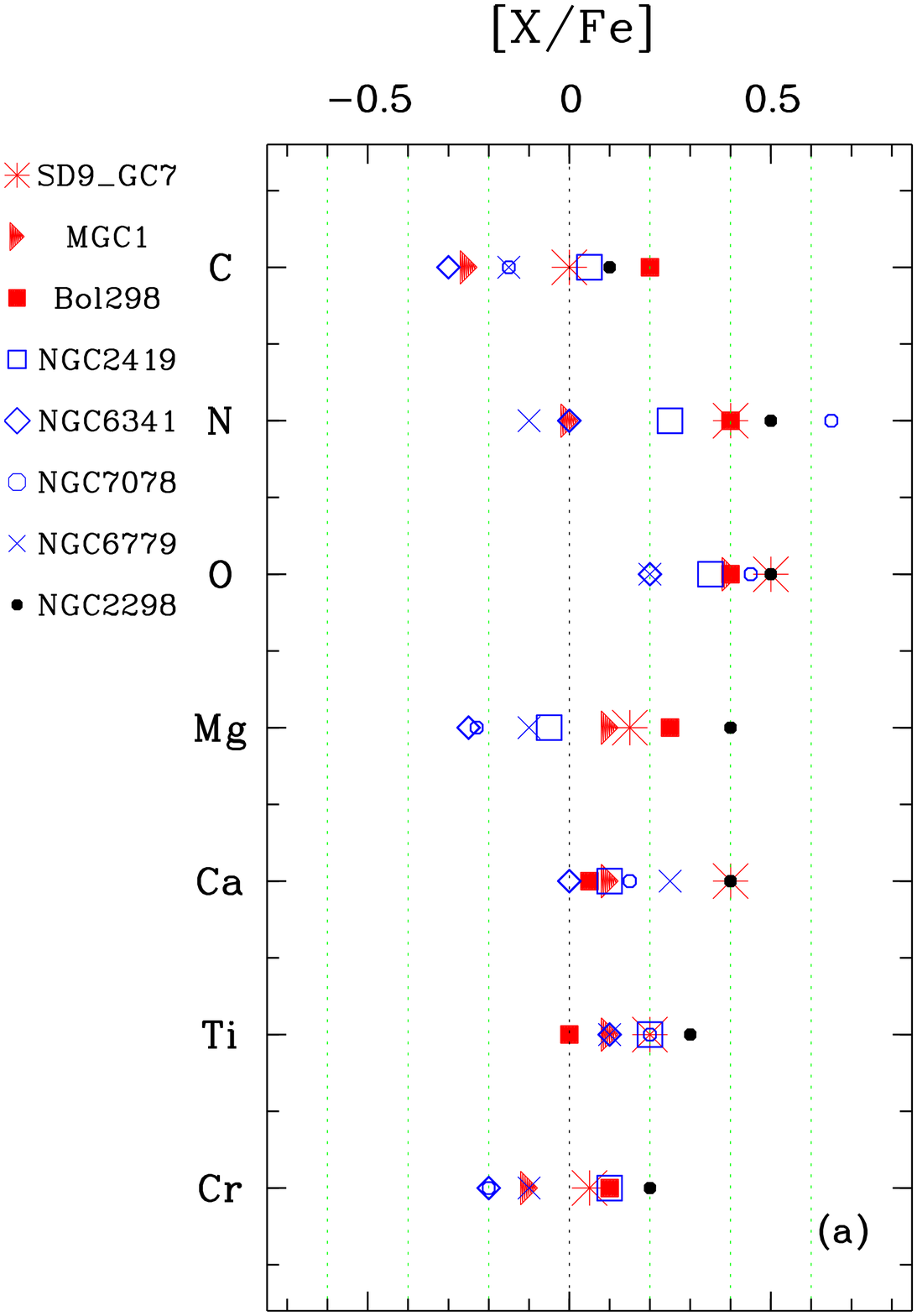} &
 \hspace{-0.4cm} 
 \includegraphics[width=1.85in,angle=-180]{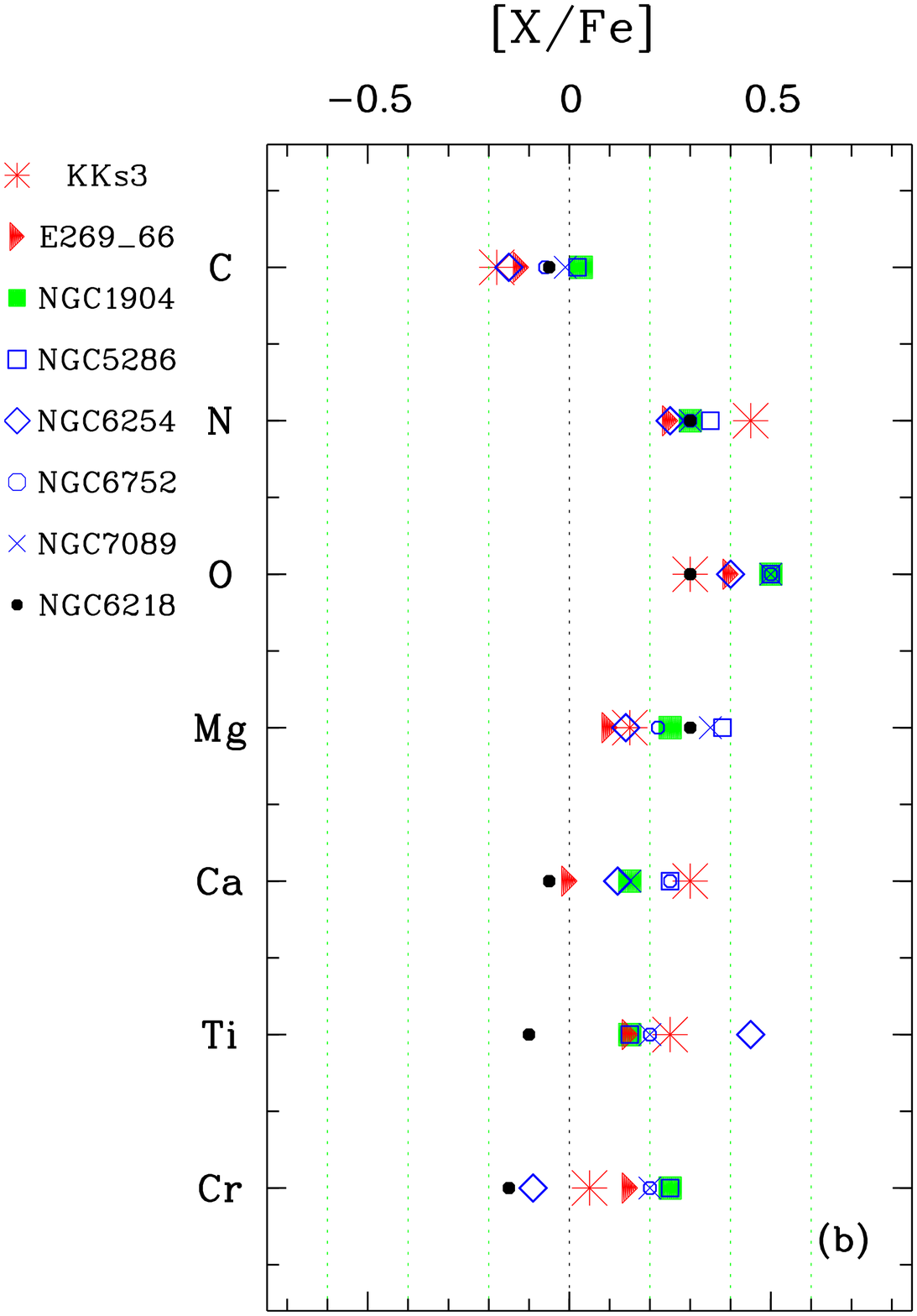} &
 \hspace{-0.4cm}
 %\hspace{-1cm}
 \includegraphics[width=1.85in,angle=-180]{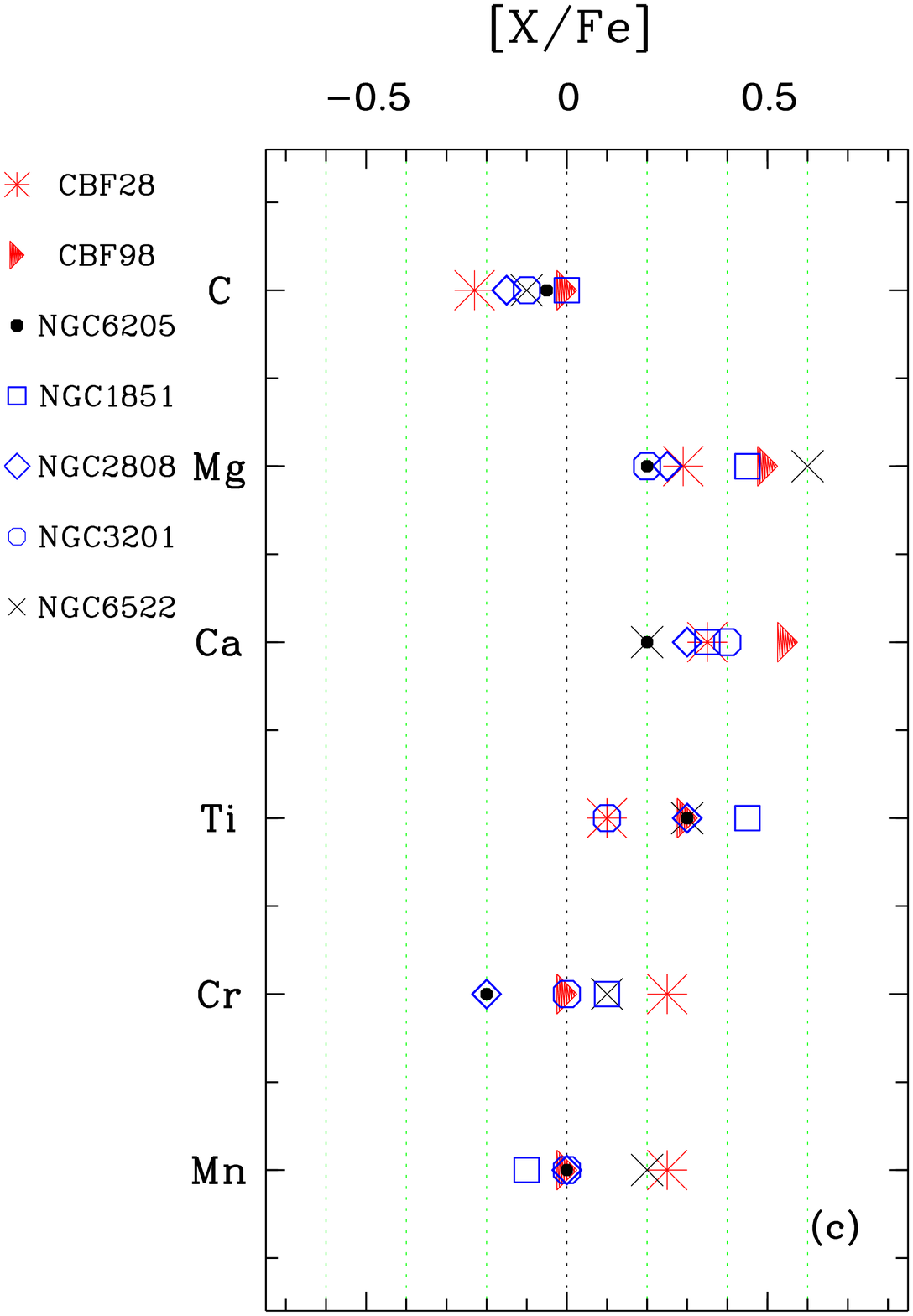} \\
\end{tabular}
 \caption{Abundances of chemical elements determined using our method for GCs in the M31 neighbourhood 
 ([SD09]~GC7, MayallII, MGC~1 and Bol~298) (panel a), for nuclear GCs in dSphs ESO269-66 and KKs3 (panel b) 
 and for GCs in M33 (CBF~98 and CBF~28) (panel c) and for Galactic GCs of similar metallicity.} 
\label{fig3}
\end{center}
\end{figure}

Table\,\ref{tab1}, Fig.\,\ref{fig2} (panel c) and Fig.\,\ref{fig3} present the results of the determination 
of ages, Y and elemental abundances for the studied extragalactic GCs. Elemental abundances for Galactic GCs of
similar metallicities are shown for comparison in Fig.\,\ref{fig2} (panel c) and Fig.\,\ref{fig3}. One can see that 
on average the chemical patterns look similar for massive extragalactic and Galactic GCs of similar metallicity.

{\it Acknowledgements.} This work is supported by the RFBR
%Russian Scientific Foundation for Basic Research 
grant No. 18-02-00167 a.


\begin{thebibliography}{}

\bibitem[Bertelli \etal\ (2008)]{Bertelli08}
{Betelli G., Girardi L., Marigo P., \& Nasi E.} 2008,
\textit{A\&A}, 484, 815 

\bibitem[Castelli \& Kurucz (2003)]{CastelliKurucz03}
{Castelli F. \& Kurucz R.L} 2003, in: Piskunov N. et al. (eds.), \textit{ Modeling of Stellar Atmospheres},
Proc. IAU Symposium No.210 (Dordrecht: Kluwer), p.\,A20

\bibitem[Chabrier (2005)]{Chabrier05}
{Chabrier G.} 2005, in: Corbelli E., Palle F., (eds.), 
\textit{ The Initial Mass Function 50 Years Later}, 
Astrophysics and Space Science Library, 327 (Berlin: Springer-Verlag), p.\,41

\bibitem[Conroy \etal\ (2018)]{Conroy18}
{Conroy C., Villaume A., van Dokkum P. G., \& Lind K.} 2018, \textit{ApJ}, 854, 139

\bibitem[Fan \etal\ (2011)]{Fan11}
{Fan Z., Huang Y.-F., Li J.-Z. et al.} 2011, \textit{Research in Astronomy and Astrophysics}, 11, 1298

%\bibitem[Gratton \etal\ 2012]{Gratton12}
%{Gratton R. G., Carretta E., Bragaglia A.} 2012, \textit{A\&AR}, 20, 50
\bibitem[Harris (1996)]{Harris96}
{Harris W.E.} 1996, \textit{AJ}, 112, 1487 (2010 edition)

\bibitem[Huchra \etal\ (1991)]{Huchra91}
{Huchra J.P., Brodie J.P., Kent S.M.} 1991, \textit{ApJ}, 370, 495

\bibitem[Kurucz (1994)]{Kurucz94}
{Kurucz R.L.} 1994, \textit{CD-Room} No. 19–22. Smithsonian Astrophysical Observatory (Cambridge)

\bibitem[Larsen \etal\ (2017)]{Larsen17}
{Larsen S. S., Brodie J. P., Strader J.} 2017, \textit{A\&A}, 601, 96

\bibitem[Ma \etal\ (2009)]{Ma09}
{Ma J. et al.} 2009, \textit{Research in Astronomy and Astrophysics}, 9, 641

\bibitem[Ma \etal\ (2012)]{Ma12}
{Ma J. et al.}, 2012, \textit{Research in Astronomy and Astrophysics}, 12, 115

\bibitem[Mackey \etal\ (2007)]{Mackey07}
{Mackey A.D., Huxor A., Ferguson A.M.N., et al.} 2007, \textit{ApJ}, 655, L85

\bibitem[Meylan \etal\ (2001)]{Meylan01}
{Meylan G., Sarajedini A., Jablonka P., et al.} 2001, \textit{AJ}, 122, 830

\bibitem[Pritzl \etal\ (2005)]{Pritzl05}
{Pritzl B.J., Venn K.A., Irwin M.} 2005, \textit{AJ}, 130, 2140

\bibitem[Roediger \etal\ (2014)]{Roediger14}
{Roediger J.C., Courteau S., Graves G., Schiavon R.P.} 2014, \textit{ApJS}, 210, 10

\bibitem[Schiavon \etal\ (2005)]{Schiavon05}
{Schiavon R.P., Rose J.A., Courteau S., MacArthur L.A.} 2005, \textit{ApJS}, 160, 163

\bibitem[Sharina \etal\ (2018)]{Sharina18}
{Sharina M.E., Shimansky V.V.} 2018, \textit{Astrophysical Bulletin}, 73, 318

\bibitem[Sharina \etal\ (2017)]{Sharina17}
{Sharina M.E., Shimansky V.V., Kniazev A.Y.} 2017, \textit{MNRAS}, 471, 1955

\bibitem[Sharina \etal\ (2009)]{Sharina09}
{Sharina M.E., Davoust E.} 2009, \textit{A\&A}, 497, 65

\bibitem[Sharina \etal\ (2010)]{Sharina10}
{Sharina M.E., Chandar R., Puzia T.H., Goudfrooij P., Davoust E.} 2010, \textit{MNRAS}, 405, 839

\bibitem[Veljanoski \etal\ (2013)]{Veljanoski13}
{Veljanoski J., Ferguson A. M. N.,  Mackey  A. D., et al.} 2013, \textit{ApJ}, 768, L33

\end{thebibliography}
\end{document}